\documentclass{iopart}

\usepackage{iopams}
\usepackage{stackrel}
\usepackage[utf8]{inputenc}
\usepackage{amssymb}
\usepackage{amsthm}
\usepackage{tikz}
\usepackage{mathrsfs}
\usepackage{epsf}
\usepackage{bookmark,hyperref}
\hypersetup{
     colorlinks=true,
     linkcolor=blue,
     filecolor=blue,
     citecolor = blue,
     urlcolor=cyan,
     }
\usepackage{pstricks}

\usepackage[perpage]{footmisc}

\newcommand{\N}{{\mathbb N}}

\newcommand{\R}{{\mathbb R}}

\newcommand{\I}{{\mathcal{O}}}

\newcommand{\grad}{\mathsf{ grad}}

\newcommand{\gi}{\mathrm{ g}}

\newtheorem{teo}{Theorem}[section]
\newtheorem{lema}[teo]{Lemma}
\newtheorem{cor}[teo]{Corollary}
\newtheorem{prop}[teo]{Proposition}

\newtheorem{rmk}{Remark}[section]

\begin{document}

\title[Strong degenerate constraining in Lagrangian dynamics]{Strong degenerate constraining in Lagrangian dynamics}

\author{J. M. Burgos}

\address{Departamento de Matem\'aticas, CINVESTAV--\,CONACYT, Av. Instituto Polit\'ecnico Nacional 2508, Col. San Pedro Zacatenco, 07360 Ciudad de M\'exico, M\'exico.}
\eads{\mailto{burgos@math.cinvestav.mx}}

\begin{abstract}
We study the strong constraining problem in Lagrangian dynamics in the degenerate codimension one case. This is the first time that degenerate potentials at the constraint are considered for this problem. Besides the total degenerate cases, these new results cover several real analytic potentials that the previous do not. Some counterintuitive effects of the degenerate constraining are discussed.
\end{abstract}


\maketitle


\section{Introduction}

The theory of constrained mechanical systems was elaborated by Lagrange in \cite{Lagrange} and it is based on a principle first stated by D'Alembert. The mentioned principle is dynamical and states, as originally stated, that for every instant the motion is such that the total work done by all the constraint's forces under any set of virtual displacements is zero. Nowadays, this principle is known as the \textit{D'Alembert-Lagrange principle} and the constraints verifying this principle are called \textit{ideal constraints}.

However, instead of molding our reality subordinating it to our mathematical theories, the right path to follow is quite the opposite. Instead of asking 
for the right definition of a constraint that gives a sufficient condition for the least-action principle to hold, the right question is whether the motion equations derived from this principle are good approximations to the real motion under the influence of real constraints. See the sections 3.6 and 3.10 in Gallavotti's book \cite{Gallavotti} for a nice discussion on ideal and real constraints as well as for historical references.

Therefore, a natural question is whether an ideal holonomic constraint is a \textit{real constraint} resulting from the limit of stiff potentials that force the motion to the constraint. Concretely, we wonder whether the motion equations resulting from the limit of stiff potentials coincide with those resulting from the least action principle subject to the respective holonomic constraints. It is very interesting and a priori counterintuitive that the answer in general is no and the responsible is the high frequency behaviour at the limit.

The paradigmatic example of a high frequency limit is the following: Consider the family of functions $\left(\cos(x/\varepsilon)\right)_{\varepsilon>0}$ on the interval $[0,1]$. By the Riemann-Lebesgue Lemma, this family weakly converges to the null function as $\varepsilon\to 0^{+}$. However, its square $\left(\cos(x/\varepsilon)^{2}\right)_{\varepsilon>0}$ weakly converges to the constant $1/2$ function.

The phenomena concerning high frequency limits is very interesting and most of the time counterintuitive. A more striking example is the Kapitza's inverted pendulum (\cite{Arnold}, section 25.E)
$$\ddot{\theta}_\varepsilon = \left( a+\frac{b}{\varepsilon}\cos\left(\frac{t}{\varepsilon}\right)\right)\sin(\theta_\varepsilon),\ \ \ \theta_\varepsilon(0)=\alpha,\ \dot{\theta}_\varepsilon(0)= \beta,\ t\geq 0,\ \varepsilon>0.$$
At the high frequency limit, the solutions $\theta_\varepsilon$ converge locally uniformly to the solution of the problem
$$\ddot{\theta}= a\sin(\theta)- \frac{b^{2}}{4}\sin(2\theta),\ \ \ \theta(0)=\alpha,\ \dot{\theta}(0)= \beta,\ t\geq 0.$$
The null equilibrium solution is Lyapunov stable, explaining the stability of the Kapitza's inverted pendulum at the high frequency limit. See \cite{Evans_av} for a proof using weak convergence methods.

The previous example is very illustrative and it shows that the effective equation for the limit solution differs from the average equation (in terms of the weak limit) that one naively would have expected. Actually, the first term of the limit equation is the average while the second is due to high frequency effects.

In view of this, we wonder what is the effective potential and the respective effective motion equation for real constraints. The first result on this subject is in the work of Rubin and Ungar in \cite{RubinUngar}. There, the constraining potentials are of the form
$$U_k(x)= \mu_k\sum_{i=1}^{s}G_i(x)^{2}$$
where $\mu_k\to+\infty$ and $(G_1,\ldots G_s)$ is regular. This result was also derived later in physical form in (\cite{KoppeJensen}, eq. 5) and (\cite{vanKampen}, eq.'s 8.33a-b) for codimension one.

In \cite{Takens} \footnote{F. Takens claimed that his ``research started with an attempt to clarify some provoking remarks in (\cite{Arnold}, sections 17.A and 21.A).".}, Takens considered the family of potentials $\lambda U$, $\lambda>0$, such that $U$ vanishes on some euclidean submanifold $M$ and is strictly positive outside it. He also asked for the second order normal derivative of $U$,
$$D^{2,\perp}_x\, U: T_x M^{\perp}\rightarrow T_x M^{\perp},\qquad \langle\, w,\,D^{2,\perp}_x\, U(v)\,\rangle=d^2_x\, U(w,v),\quad w,v\in T_x M^{\perp},$$
to be smoothly diagonalizable along $M$; that is
$$D^{2,\perp}_x\, U= \sum_{i=1}^{k}\sigma_i(x)\ P_i(x),$$
for every $x$ in $M$ such that every $\sigma_i(x)$ is nonnegative and the following is a complete set of orthogonal idempotent elements
$$\sum_{i=1}^{k}P_i(x)= id\,|_{T_x M^{\perp}},\quad P_i^{2}=P_i,\quad P_iP_j={\bf 0},\ i\neq j$$
where each projection $P_i$ is smooth along $M$.
He calculated the effective motion equation and showed that if the \textit{non-resonant condition}
$$\pm\sigma_i(x)^{1/2}\pm\sigma_j(x)^{1/2}\pm\sigma_k(x)^{1/2}\neq 0,\qquad \sigma_i(x)\neq \sigma_j(x),\ i\neq j$$
holds for every $x$ in $M$, then its solution is the limit as $\lambda\to 0^{+}$ of the sequence of solutions of the respective equations whose potential term is $\lambda U$. He also showed that the condition on the second normal derivative is necessary giving an example of a potential whose minimum is reached at a codimension two submanifold such that the limit set of the sequence of solutions is a funnel and in particular there is no unique limit motion. This is known as \textit{Takens chaos}.

In \cite{BS}, Bornemann and Sch\"utte used weak convergence methods to study the problem for Hamiltonian systems in codimension one. In \cite{Bornemann}, Bornemann generalized the previous result to arbitrary codimension and generalized the non-resonant condition as well.
All of the previous results in the literature assume that the potential is non-degenerate at the constraint, that is to say
$$\ker D^{2}_x\, U = T_x M,\qquad \langle\, w,\,D^2_x\, U(v)\,\rangle=d^2_x\, U(w,v),\quad w,v\in T_x M.$$


In this paper, we treat the codimension one problem for mechanical Lagrangian systems with an arbitrary degeneracy of the potential at the constraint hypersurface. This is the first time that a degenerate potential at the constraint is considered for this problem.

Concretely, consider a smooth manifold $N$ with dimension greater than one and the family of mechanical Lagrangians
\begin{equation}\label{Lagrangian}
L_\varepsilon(x,v)= K_x(v)-\varepsilon^{-2} U(x),\qquad (x,v)\in TN,\ \varepsilon>0
\end{equation}
where $K_x$ is a positive definite quadratic form for every $x$ in $N$ and $C^{2}$ with respect to $x$ and the potential $U:N\rightarrow\R$ verifies:

\bigskip
\noindent\textbf{Hypothesis:} The potential $U$ is the composition $g\circ f$ such that
\begin{enumerate}
\item $f:N\rightarrow\R$ is $C^{3}$ and zero is a regular value of $f$.
\item $g:\R\rightarrow\R$ is a nonnegative $C^{2}$ function vanishing only at zero and there is an open interval $I$ containing zero such that $e= (g/g')|_{I-\{0\}}$ has a removable singularity at the origin and removing it by defining $e(0)=0$ gives a $C^{1}$ function on $I$.
\end{enumerate}
\bigskip

The following remarks are immediate from the hypothesis and useful for the structure and understanding of the problem:
\begin{itemize}
\item There is a unique $C^{2}$ Riemannian metric $\rho$ on $N$ such that $K_x(v)=\Vert v \Vert_x^{2}/2$ for every $(x,v)$ in $TN$, where $\Vert\cdot\Vert_x$ is the norm induced by the inner product $\rho_x$ on $T_x N$. 
\item The hypothesis on $f$ implies that the global minimum of the potential, the locus
$$M=[f=0]=[U=0],$$
is a hypersurface in $N$ with dimension greater than zero and $M\times\{{\bf 0}\}$ consists entirely of equilibrium points of the Lagrangian dynamics of \eref{Lagrangian}.
\item The hypothesis on $g$ includes all of the nonnegative real analytic functions vanishing only at zero as well as functions vanishing at zero of the form
\begin{equation}\label{Infinite_zero}
g(x)= a\exp(-b x^{-2m}),\qquad a,\ b>0,\ m\in\N
\end{equation}
outside zero. 
\item Denote by $\alpha$ the derivative at zero of the function resulting by removing the singularity of $g/g'$. To gain some intuition over this parameter, as an example consider a real analytic function $g(x)= ax^{m}+o(x^{m})$ and note that the corresponding parameter is $\alpha= m^{-1}$. As another example, consider \eref{Infinite_zero} and see that $\alpha=0$ in this case.

In this sense, the parameter $\alpha$ measures how close $g$ is from zero near the origin and it does not depend on the amplitude of the function but only on its shape near the origin; i.e. constant multiples of $g$ give the same parameter. Later on, another characterization will be given in terms of the \textit{weak virial Theorem} \eref{weak_virial_intro}.
\end{itemize}


Consider a point $p$ in $M$ and a vector $v\neq{\bf 0}$ in $T_p N$. For every $\varepsilon>0$, consider the solution $x_{\varepsilon}$ of the Euler-Lagrange equations of \eref{Lagrangian} such that $x_{\varepsilon}(0)=p$ and $\dot{x}_{\varepsilon}(0)= v$. These solutions are defined over the whole real line and we are interested in their limit as $\varepsilon\to 0^{+}$.

Along the text, the gradient vector field of $f$ will always be taken with respect to the metric $\rho$ and it will be denoted by $\grad_\rho f$. We will denote by $Crit(f)$ the set of critical points of $f$.

We define the \textit{equipotential distortion} $\kappa$ at a point in $N- Crit(f)$ as the curvature of the respective gradient flow line at the point. It is a vector field tangent to the equipotential foliation on $N-Crit(f)$ and it is intrinsic to the foliation in the sense that it only depends on the class of functions $\lambda\circ f$ where $\lambda:\R\rightarrow\R$ is $C^{1}$ with non vanishing derivative at every point. In particular, given an equipotential foliation, the concept of equipotential distortion of a given leaf is well defined.

As an example, the equipotentials of a radial function are homothetic and their equipotential distortion is zero. In this sense, $\kappa$ measures the shape distortion of the equipotential hypersurfaces.

A more interesting example is provided by the potential
\begin{equation}\label{example1}
U(x)=(x^{2}+2y^{2}+3z^{2}-1)^{4}
\end{equation}
constraining at an ellipsoid whose equipotential distortion is a vector field with critical points in the axes. The critical points lying in the $x$-axis are repeller points, those in the $y$-axis are saddle points and the ones in the $z$-axis are attractors.

With respect to the metric $\rho$, there is a unique orthogonal splitting of the initial velocity $v= v_{\Vert}+v_{\perp}$ such that $v_{\Vert}$ is in $T_p M$ and $v_{\perp}$ is in $(T_p M)^{\perp}$. Recall that $\alpha$ is the derivative at zero of the function resulting by removing the singularity of $g/g'$ at zero.

\begin{teo}\label{Main}
The family $(x_\varepsilon)_{\varepsilon>0}$ converges locally uniformly to a $C^{2}$-smooth curve $x:\R\rightarrow M$ which is the unique solution of
\begin{equation*}
\fl \qquad\nabla_{\dot{x}}\,\dot{x}\, +\,\frac{\Vert v_{\perp}\Vert_p^{2}}{2\alpha+1}\, \left(\frac{\Vert\grad_\rho f(x)\Vert_x}{\Vert\grad_\rho f(p)\Vert_p}\right)^{\frac{2}{2\alpha +1}}\,\kappa(x)=0,\qquad x(0)=p,\quad \dot{x}(0)= v_\Vert
\end{equation*}
where $\nabla$ denotes the Levi-Civita connection with respect to the induced ambient metric on $M$.
\end{teo}

Note that, because the solution is unique, if the family of Lagrangians \eref{Lagrangian} is $C^{\infty}$-smooth, then the equation will be so hence its solution $x$ will be $C^{\infty}$ as well.

The effective force field due to high frequency effects is either zero or collinear with and opposite to the equipotential distortion on $M$. If the equipotential distortion is non null at $p$, even if $v_\Vert={\bf 0}$ there is a non null acceleration at $p$. It is also interesting and counterintuitive that the effective force does not depend on the amplitude of $g$ but only of its shape near zero, i.e. constant multiples of $g$ give the same effective force field.

The effective force field is conservative and it comes from an effective potential $U_{eff}$ described below.

\begin{cor}\label{Main_Cor_1}
The dynamics of the limit motion is governed by the effective mechanical Lagrangian $L_{eff}(x,v)= K_x(v)-U_{eff}(x)$ on $TM$ where
$$U_{eff}(x)\,=\, \theta\, (\alpha+1/2)\, \Vert\grad_\rho f(x)\Vert_x^{\frac{2}{2\alpha+1}}\,=\,\frac{\Vert v_{\perp}\Vert_p^{2}}{2}\, \left(\frac{\Vert\grad_\rho f(x)\Vert_x}{\Vert\grad_\rho f(p)\Vert_p}\right)^{\frac{2}{2\alpha +1}}$$
up to an additive constant and $\theta$ is the \textit{adiabatic invariant}
$$\theta\,=\,\frac{1}{2\alpha+1}\, \Vert v_{\perp}\Vert_p^{2}\ \Vert\grad_\rho f(p)\Vert_p^{-\frac{2}{2\alpha+ 1}}.$$
\end{cor}

It is interesting that while the effective potential is linear with respect to the gradient's norm in the non-degenerate case ($\alpha=1/2$), in the infinite degenerate case ($\alpha=0$) it is quadratic.

With respect to the gradient flow lines of $f$, the transverse kinetic and potential energy in the limit, $T_\perp$ and $U_\perp$ respectively, verify the \textit{weak virial Theorem}
\begin{equation}\label{weak_virial_intro}
2\alpha\ T_\perp= U_\perp.
\end{equation}
At the weak limit as $\varepsilon\to 0$, this is the generalization of the Bornemann's version (\cite{Bornemann}, section 2.6) of the virial Theorem (\cite{AbrahamMarsden}, Theorem 3.7.30) for non-degenerate potentials and named by him as the weak virial Theorem. In particular, in the non-degenerate case there is an equipartition between the transverse kinetic and potential energy while in the infinite degenerate case, all of the transversal energy is kinetic. This partition and the existence of an adiabatic invariant for the transverse motion are the basis for the existence of the effective potential.

Given a non tangential initial condition at the constraint, very interesting behaviour can occur. As an example, consider the following potential in Newtonian dynamics
$$U(x,y)= \left(y\ e^{x^{2}}\right)^{2m},\qquad m\in\N.$$
This potential constrains at the $x$-axis and for every non tangential initial condition on the axis, there is a strictly positive constant $C$ such that the effective potential there is
$$U_{eff}(x)= C\ \left(e^{x^{2}}\right)^{\frac{2m}{m+1}}.$$
In particular, because of the effective potential's convexity, the limit curve is trapped in the respective Hill's region. The approximated curves for small $\varepsilon$ are like Lissajous curves going back and forth and oscillating across the axis for a long period of time. This analogous to what happens in the magnetic mirror but here the effect is purely mechanical.

\begin{cor}\label{Main_Cor_2}
If $v$ is in $T_p M$, then the family $(x_\varepsilon)_{\varepsilon>0}$ converges locally uniformly to a geodesic in $M$ with initial conditions $x(0)=p$ and $\dot{x}(0)=v$ with respect to the induced ambient metric on $M$.
\end{cor}

The interesting and new fact about the previous result is that now, for codimension one, it is valid in a wider class of potentials besides those non-degenerate at the constraint.

As an example, consider again the potential \eref{example1}. Given a tangential initial condition, the respective family of curves converge locally uniformly to the respective geodesic on the ellipsoid. Because the potential is degenerate at the ellipsoid, non of the previous results in the literature apply.

\begin{cor}\label{Main_Cor_3}
The hypersurface $M$ has null equipotential distortion iff, for every point $p$ in $M$ and vector $v$ in $T_p N$ with orthogonal projection $v_\Vert\neq{\bf 0}$ in $T_p M$, the family $(x_\varepsilon)_{\varepsilon>0}$ converges locally uniformly to a geodesic in $M$ with initial conditions $x(0)=p$ and $\dot{x}(0)=v_{\Vert}$ with respect to the induced ambient metric on $M$.
\end{cor}

The Lagrangian on $M$ considered as an ideal constraint is just the Lagrangian \eref{Lagrangian} restricted to $TM$ and because it is only kinetic, its dynamics consist of geodesic motions in $M$ with respect to the induced ambient metric. On the contrary, the Lagrangian on $M$ considered now as a real constraint is the effective Lagrangian described in Corollary \ref{Main_Cor_1} whose dynamics are the solutions of the equation in the main theorem.

Therefore, an ideal constraint is real if and only if it has null equipotential distortion. In particular, the condition for an ideal constraint to be real depends on the equipotential foliation geometry near the constraint and not only on the intrinsic geometry of it.

The direct implication of  corollary \ref{Main_Cor_3} is a particular case of a general result in arbitrary codimension (\cite{Gallavotti}, section 3.8, Proposition 13, ``Arnold's Theorem"). See also (\cite{Arnold}, section 21.C). The interesting thing about the previous corollary is the converse: even for an infinite degenerate potential at the constraint, the vanishing of the equipotential distortion is necessary in order for the the constraint to be real. The converse for non-degenerate potentials at the constraint was proved in (\cite{Bornemann}, section 3.2, Theorem 3). Again, in view of the previous remark, corollary \ref{Main_Cor_3} is a priori counterintuitive for one would have expected that for an infinite degenerate potential, the ideal constraint would be real no matter what.

\section{Preliminaries on weak convergence}

This section concerns some weak convergence preliminaries. The presentation will be minimalistic and specially suited for our needs. We refer the reader to the classical functional analysis reference \cite{Rudin} and to \cite{Adams} for Sobolev spaces.

Consider a real Banach space $(E,\Vert\cdot\Vert_E)$ and its dual space $E^{*}$ consisting of bounded linear functionals of $E$. The operator norm $\Vert\cdot\Vert_{E^{*}}$ induces the \textit{strong topology} on $E^{*}$. The \textit{weak topoloqy} $\omega$ on $E^{*}$ is the coarsest topology such that every functional on $E^{**}$ is continuous. In particular,
$$f_n\mathop{\rightharpoonup}^{\omega} f\qquad if\qquad F(f_n)\rightarrow F(f),\ F\in E^{**}.$$
There is a canonical isometric embedding of the space $E$ on its bidual space
$$J:E\rightarrow E^{**},\qquad (x\mapsto \hat{x}),\qquad \hat{x}(h)= h(x).$$

The \textit{weak star topology} $\omega^{*}$ on $E^{*}$ is the coarsest topology such that every functional in $J(E)$ is continuous. In particular,
$$f_n\mathop{\rightharpoonup}^{\omega^{*}} f\qquad if \qquad f_n(h)\rightarrow f(h),\ h\in E.$$
If the space $E$ is reflexive, i.e if $J$ is an isomorphism, then the weak and weak star topologies on $E^{*}$ coincide. Moreover, the converse is also true.

As a direct application of the Banach-Steinhaus Theorem we have:

\begin{prop}\label{Bound}
Every weakly star convergent sequence in $E^{*}$ is bounded.
\end{prop}

\noindent As a direct application of the Banach-Alaoglu Theorem we have:

\begin{prop}\label{Alaoglu}
If $E$ is separable, then every bounded sequence in $E^{*}$ has a weakly star convergent subsequence.
\end{prop}

Now we specialize in the $L^{p}$ spaces over an interval $I= (-T,T)$ with $T>0$. Consider $p,\ q> 1$ such that $p^{-1}+q^{-1}=1$ and recall that
$$L^{p}(I)\cong L^{q}(I)^{*},\qquad L^{\infty}(I)\cong L^{1}(I)^{*}$$
where the evaluation is by integration
$$x(h)= \int_{-T}^{T}\,dt\ x(t)h(t).$$
In particular, the $L^{p}(I)$ spaces with $1<p\leq \infty$ have a weak star topology such that the closed unit ball is sequentially compact. Because these spaces are reflexive for $1<p<\infty$, the weak and weak star topologies coincide in these cases. From now on we will exclusively work in the weak star topology.

The space $L^{\infty}(I)$ is a Banach algebra, i.e. $\Vert ab\Vert_\infty\leq \Vert a\Vert_\infty \Vert b\Vert_\infty$, and the space of continuous functions $C(\bar{I}\,)$ is a Banach subalgebra.

Because of H\"older's inequality, every $L^{p}(I)$ space is an $L^{\infty}(I)$--module. Every $L^{p}(I)$ space with the weak star topology is also an $L^{\infty}(I)$--module as the next proposition shows.

\begin{prop}\label{module}
If $y_n\rightarrow y$ strongly in $L^{\infty}(I)$ and $x_n\rightharpoonup x$ weakly star in $L^{p}(I)$, then $y_n x_n\rightharpoonup y x$ weakly star in $L^{p}(I)$.
\end{prop}
\begin{proof}
For every test function $h$ in $L^{q}(I)$ we have
$$y_n x_n(h)- yx(h)= (y_n-y)x_n (h) + (x_n-x)y(h)$$
$$= (y_n-y)x_n (h) + (x_n-x)(yh)\rightarrow 0$$
for the first term is bounded by H\"older's inequality
$$\vert(y_n-y)x_n (h)\vert\leq \Vert y_n-y \Vert_\infty \Vert x_n\Vert_p \Vert h \Vert_q\rightarrow 0$$
and the second term goes to zero as well.
\end{proof}

A function $x$ in $L^{p}(I)$ has a \textit{weak derivative} in $L^{p}(I)$ if there is some $y$ in $L^{p}(I)$ such that
$$y(h)= -x(\dot{h}),\qquad \forall\,h\in C_c^{1}(I).$$
where $C_c^{1}(I)$ is the space of differentiable real valued functions with compact support on $I$. In this case one sets $\dot{x}:= y$. If a weak derivative exists, then it is unique.

Define the \textit{Sobolev space} $W^{1,p}(I)$ as the linear subspace of $L^{p}(I)$ whose elements have weak derivative in $L^{p}(I)$ and Sobolev norm
$$\Vert x\Vert_{W^{1,p}}= \Vert x \Vert_p +\Vert \dot{x}\Vert_p.$$
In the case where $p=2$, the Sobolev space is denoted by $H^{1}(I)$ and the norm is induced by the inner product
$$\langle\, x,\, y\,\rangle_{H^{1}}= \langle\, x,\, y\,\rangle_2 + \langle\, \dot{x},\, \dot{y}\,\rangle_2.$$

\begin{prop}
The Sobolev space $W^{1,p}(I)$ is identified with a closed linear subspace of $L^{p}(I)\times L^{p}(I)$ with the strong topology under the map $x\mapsto (x,\dot{x})$. In particular, it is a Banach space.
\end{prop}
\begin{proof}
It is clear that the image of the map is a linear subspace. It rests to show the other assertion.

Consider an accumulation point $(x,y)$ of the image of the map. The product space is a metric space, in particular first countable hence there is a sequence $(x_n)$ in $W^{1,p}(I)$ such that $x_n\to x$ and $\dot{x}_n\to y$ as $n\to +\infty$ strongly in $L^{p}(I)$. Then,
$$y(h)\leftarrow \dot{x}_n(h)= -x_n(\dot{h})\rightarrow -x(\dot{h}),\qquad h\in C_c^{1}(I)$$
and we conclude that $\dot{x}= y$ for the limit and the weak derivative are unique. In particular, $(x,y)=(x,\dot{x})$ belong to the image of the map and because the choice of the point was arbitrary, we have proved that the image is closed.
\end{proof}

Identifying $W^{1,p}(I)$ with this closed linear subspace, the weak star topology on this space is defined as the subspace topology induced by the weak star topology of the product. In particular, because the weak star topology of the product is the product topology of the weak star topologies of the respective factors, we have that  $x_n\rightharpoonup x$ weakly star on $W^{1,p}(I)$ if and only if $x_n\rightharpoonup x$ and $\dot{x}_n\rightharpoonup \dot{x}$ weakly star on $L^{p}(I)$.

\begin{prop}\label{Alaoglu_W}
If $1<p\leq \infty$, then every bounded sequence in $W^{1,p}(I)$ has a weakly star convergent subsequence.
\end{prop}
\begin{proof}
If $\Vert x_n\Vert_{W^{1,p}}\leq M$ for some $M>0$ and every natural $n$, then $\Vert x_n\Vert_{p},\ \Vert \dot{x}_n\Vert_{p}\leq M$ for every natural $n$. By Proposition \ref{Alaoglu}, there is a subsequence $(n_i)$ such that $x_{n_i}\rightharpoonup x$ and $\dot{x}_{n_i}\rightharpoonup y$ weakly star on $L^{p}(I)$. Then, 
$$y(h)\leftarrow \dot{x}_n(h)= -x_n(\dot{h})\rightarrow -x(\dot{h}),\qquad h\in C_c^{1}(I)$$
and we conclude that $\dot{x}= y$ for the limit and the weak derivative are unique. By the previous remark, we conclude that $x_{n_i}\rightharpoonup x$ weakly star on $W^{1,p}(I)$.
\end{proof}

\begin{prop}\label{magic}
Suppose that $1<p\leq \infty$ and consider a sequence $(x_n)$ in $W^{1,p}(I)$ such that $x_n\rightharpoonup {\bf 0}$ weakly star in $L^{p}(I)$ and the sequence of weak derivatives is uniformly bounded in $L^{p}(I)$. Then, $\dot{x}_n\rightharpoonup {\bf 0}$ weakly star in $L^{p}(I)$.
\end{prop}
\begin{proof}
There is $M>0$ such that $\dot{x}_n$ is in $B= \overline{B({\bf 0}, M)}$ for every natural $n$. The weakly star compact space $B$ is metrizable by a metric $d$.

Suppose that $(\dot{x}_n)$ does not weakly star converge to ${\bf 0}$. Then, there is $\varepsilon>0$ and a subsequence $(n_i)$ such that
$$d(\dot{x}_{n_i},{\bf 0})\geq \varepsilon$$
for every $i$. By Proposition \ref{Bound}, $(x_{n_i})$ is bounded in $W^{1,p}(I)$ and by Proposition \ref{Alaoglu_W} there is a subsequence $(i_j)$ such that $x_{n_{i_j}}\rightharpoonup x$ weakly star on $W^{1,p}(I)$ for some $x$ in $W^{1,p}(I)$. In particular, $x_{n_{i_j}}\rightharpoonup x$ and $\dot{x}_{n_{i_j}}\rightharpoonup \dot{x}$ weakly star on $L^{p}(I)$. Because of the limit uniqueness, $x=0$ therefore $\dot{x}=0$ and we conclude that
$$\dot{x}_{n_{i_j}}\rightharpoonup {\bf 0}$$
weakly star on $L^{p}(I)$, which is absurd.
\end{proof}

Consider the Cauchy problem given by the ordinary differential equation and initial condition
\begin{equation}\label{Cauchy_problem_prel}
\dot{x}= F(t,x),\qquad x(0)= x_0
\end{equation}
such that $F$ is continuous and locally Lipschitz in the second variable. A \textit{weak solution} of the problem above is a solution $x$ in $W^{1,p}(I)$ of the integral equation
\begin{equation}\label{Cauchy_problem_prel_int}
x(t)= x_0+\int_0^{t}\, ds\ F(s,x(s)).
\end{equation}
Every differentiable solution in $C^{1}(\bar{I})$ of the problem \eref{Cauchy_problem_prel} is called a \textit{strong solution}.

\begin{prop}\label{weak_strong}
Every weak solution of \eref{Cauchy_problem_prel} is strong. In particular, it is unique.
\end{prop}
\begin{proof}
For every $1\leq p\leq \infty$, we have the Sobolev embedding\footnote{Actually, $W^{1,1}(I)$ constitutes the absolute continuous functions while $W^{1,\infty}(I)$ is the set of Lipschitz continuous functions.}
$$W^{1,p}(I)\subset C(\bar{I})$$
and by the local existence and uniqueness Picard's Theorem, there is a unique solution of \eref{Cauchy_problem_prel_int} among the continuous functions and this solution is $C^{1}$ regular.
\end{proof}

\section{Preliminaries on the limit curve and suitable coordinates}\label{Subsection_limit_curve}

Recall from the introduction that $p$ is a point in $M$, $v\neq{\bf 0}$ is a vector in $T_p N$ and for every $\varepsilon>0$, $x_{\varepsilon}$ is the solution of the Euler-Lagrange equations of \eref{Lagrangian} defined on the respective maximal interval with initial conditions $x_{\varepsilon}(0)=p$ and $\dot{x}_{\varepsilon}(0)= v$.

The following result is the obvious adaptation of the one proved in (\cite{BMP}, page 4) and was proved in (\cite{BP}, pages 4339-4340) in the context of Lagrangian mechanics or equivalently with a Riemannian metric on $N$ induced by the kinetic term. The proof consists of an energy argument.

\begin{lema}\label{lema_preliminaries}
For every $\varepsilon>0$, the curve $x_{\varepsilon}$ verifies:
\begin{enumerate}
\item It is defined on the whole real line.

\item\label{item2_lema_prel} For every $t$ in the real line the velocity is bounded and the bound does not depend on $\varepsilon$,
$$\Vert\dot{x}_\varepsilon(t)\Vert_{x_{\varepsilon}(t)}\leq \Vert v \Vert_p,.$$
Moreover, the image of the curve is contained in a neighbourhood of $M$,
$${\rm Im}(x_\varepsilon)\subset [U\leq \varepsilon^{2}\,\Vert v \Vert_p^{2}/2].$$

\item For every $T>0$, the segment of the curve in the interval $[-T,T]$ verifies
$${\rm Im}\left(x_{\varepsilon}|_{[-T,T]}\right)(t)\subset\overline{B_\rho(p, T\, \Vert v \Vert_p)}$$
where the ball is with respect the distance induced by the metric $\rho$. Note that the region is a compact set not depending on $\varepsilon$.
\end{enumerate}
\end{lema}

\begin{cor}\label{Cor3}
Let $T>0$. There is a continuous curve $x:[-T,T]\rightarrow M$ with $x(0)=p$ and a sequence $(\varepsilon_j)$ such that $\varepsilon_j>0$, $\varepsilon_j\to 0^{+}$ and $x_{\varepsilon_j}\rightarrow x$ uniformly on $[-T,T]$.
\end{cor}

Note that in the second item of Lemma \ref{lema_preliminaries}, while the velocity is bounded by a constant independent of $\varepsilon$, the neighbourhood containing the curve shrink to $M$ as $\varepsilon\to 0^+$. In particular, the sequence in Corollary \ref{Cor3} will not converge in the $C^1$-topology in general. As an example, consider the family of potentials
$$U_n(x)\,=\, n^2\,x^2\,/\,2,\qquad x\in\R$$
in the context of Newtonian dynamics in the real line. The resulting sequence of motions $x_n$ with $x_n(0)=0$ and $\dot{x}_n(0)=v_0\neq 0$ uniformly converges to zero but does not converge in the $C^1$-topology.

In view of the lack of $C^1$-convergence in general, there is no a priori reason to expect the limit curve $x$ in Corollary \ref{Cor3} to be smooth. However, we will prove in the next section that our hypothesis described at the introduction are sufficient to guarantee that this limit will be at least $C^2$. For this purpose, we construct the following suitable coordinates for the problem.

Consider the flow $\phi$ in $N- Crit(f)$ of the Cauchy problem
\begin{equation}\label{Cauchy3}
\partial_t \phi= \frac{\grad_\rho f}{\Vert \grad_\rho f \Vert_\phi^{2}}(\phi),\ \phi(0,x)= x,\ x\in N-Crit(f).
\end{equation}

Consider an arbitrary $T>0$ and an arbitrary point $q$ in $M$. Consider a local coordinate neighbourhood $(V, \psi)$ of $M$ centered at $q$. By hypothesis, $M$ is contained in $N-Crit(f)$. We define the $C^{2}$ map $\Psi$ by the expression
$$\Psi(r,y)= \phi(r, \psi(y)).$$

The following Lemma is proved in (\cite{BMP}, Lemma 2.5) and (\cite{BP}, Lemma 2.5).
\begin{lema}\label{Lema2}
\begin{enumerate}
\item $\Psi(0,y)= \psi(y)$ for every $y$ in $\psi^{-1}(V)$.
\item $f(\Psi(r,y))= r$  and $U(\Psi(r,y))= g(r)$ for every $(r,y)$ in $\Psi^{-1}(\I(V))$.
\item $(\I(V), \Psi)$ is a local coordinate neighbourhood of $N$ centered at $q$ where
$$\I(V)\subset N-Crit(f)$$
is the union of the set of orbits of \eref{Cauchy3} with initial condition in $V$.
\end{enumerate}
\end{lema}


\section{The proof}\label{Section_proof}

\subsection{A suitable subsequence}

From now on, consider an arbitrary $T>0$. By Lemma \ref{Lema2}, for every local coordinate neighbourhood $(V, \psi)$ of $M$ we automatically have the corresponding local coordinate neighbourhood $(\I(V), \Psi)$ of $N$.

The main purpose of this subsection is to find a suitable subsequence of the one in Corollary \ref{Cor3} defining the limit curve. In particular we will prove that the limit curve is $C^1$. The strategy is to cover the image of the limit curve with local coordinate neighbourhoods of $M$ and find a subsequence of $(x_{\varepsilon_j})$ such that restricted to a segment contained in one of these neighbourhoods, the tangential coordinates of the resulting sequence converges in the $C^1$ topology while the radial coordinate goes to zero uniformly. In contrast to the tangential coordinates, the radial velocity weakly converges to zero but the convergence is not in the strong sense. The weak convergence of the radial velocity is proved in Lemma \ref{Lema4} and the fact that the convergence is not strong is a consequence of Lemma \ref{Lema7}.

In particular, we prove that the limit curve is $C^1$ and the convergence of the obtained subsequence is $C^1$ in the respective tangential coordinates and uniform in the global radial coordinate such that the radial velocity weakly converges. This subsequence is the most important object of this subsection and will be used in the following ones.

\begin{lema}\label{Lema3}
Consider $\tau$ in $[-T, T]$ and a local coordinate neighbourhood $(V, \psi)$ of $M$ centered at $x(\tau)$. Then, there is $\delta>0$ and a natural $N$ such that ${\rm Im}\left(x_{\varepsilon_j}|_{\bar{I}}\right)$ and ${\rm Im}\left(x|_{\bar{I}\cap [-T,T]}\right)$ are contained in a compact set of $\I(V)$ for every $j\geq N$ where $I=(\tau-\delta, \tau+\delta)$.
\end{lema}
\begin{proof}
Let $R>0$ be such that $\overline{B_\rho(x(\tau),R)}\subset\I(V)$. There is $\delta>0$ small enough such that $\delta\Vert v\Vert_p<R/2$ and a natural $N$ such that $d_\rho( x_{\varepsilon_j}(\tau), x(\tau))<R/2$ if $j\geq N$. Then,
$$d_\rho(x_{\varepsilon_j}(t), x(\tau))\leq d_\rho(x_{\varepsilon_j}(t), x_{\varepsilon_j}(\tau)) + d_\rho(x_{\varepsilon_j}(\tau), x(\tau))$$
$$\leq  \left\vert\int_{\tau}^{t}ds\ \Vert\dot{x}_{\varepsilon_j}(s)\Vert\right\vert +R/2\leq \vert t-\tau\vert\,\Vert v\Vert_p + R/2$$
$$\leq \delta\Vert v\Vert_p + R/2\leq R$$
for every $t$ in $\bar{I}$ and every $j\geq N$. Taking the limit $j\to+\infty$ in the previous expression, by continuity we also have
$$d_\rho\left(x(t), x(\tau)\right)\leq R$$
for every $t$ in $\bar{I}\cap [-T,T]$. The lemma is proved.
\end{proof}

\begin{prop}\label{compact_sets}
There are naturals $N_0$ and $l$ such that for every $i=1,2,\ldots l$ there is a point $\tau_i$ in $[-T,T]$, a coordinate chart $(\I(V_i), \Psi_i)$ centered at $x(\tau_i)$, a compact set $C_i$ of $\I(V_i)$ and an open interval $I_i$ centered at $\tau_i$ of the real line with the following properties:
\begin{enumerate}
\item $\tau_1<\tau_2<\ldots <\tau_l$.
\item ${\rm Im}\left(x_{\varepsilon_j}|_{\bar{I}_i}\right)$ and ${\rm Im}\left(x|_{\bar{I}_i\cap [-T,T]}\right)$ are contained in $C_i$ for every $j\geq N_0$.
\item The open intervals $I_1, I_2,\ldots, I_l$ cover $[-T, T]$.
\end{enumerate}
\end{prop}
\begin{proof}
Because the choice of $\tau$ in $[-T,T]$ was arbitrary, by Lemmas \ref{Lema2} and \ref{Lema3} we have that for every $\tau$ in $[-T,T]$, there is a coordinate chart $(\I(V), \Psi)$ centered at $x(\tau)$, an open interval $I$ centered at $\tau$ and a natural $N$ such that ${\rm Im}\left(x_{\varepsilon_j}|_{\bar{I}}\right)$ and ${\rm Im}\left(x|_{\bar{I}\cap [-T,T]}\right)$ are contained in a compact set $C$ of $\I(V)$ for every $j\geq N$.

These open intervals cover the compact interval $[-T, T]$ hence there is a finite collection $\tau_1<\tau_2<\ldots <\tau_l$ of points in $[-T, T]$ whose associated intervals $I_1, I_2,\ldots, I_l$ cover $[-T, T]$ and a natural $N_0$ that is the maximum of the associated naturals $N_1,N_2,\ldots, N_l$. Denote the respective coordinate neighbourhoods by $(\I(V_i), \Psi_i)$ and the respective compact sets by $C_i$ where $i=1,2,\ldots l$. The proof is complete.
\end{proof}

The index $i=1,2,\ldots l$ will be called the \textit{chart index}. Now, for every $j\geq N_0$ and every chart index $i$, the coordinates of the curves $x_{\varepsilon_j}$ and $x$ are defined on $\bar{I}_i$ and $\bar{I}_i\cap [-T, T]$ respectively. Concretely, the radial and tangential coordinates of the curves are defined as follows:
$$\Psi_i(r_{\varepsilon_j}(t), y_{i,\varepsilon_j}(t))=x_{\varepsilon_j}(t),\quad t\in \bar{I}_i,$$
$$\Psi_i(r(t), y_i(t))=x(t),\quad t\in \bar{I}_i\cap [-T, T].$$

\begin{rmk}\label{globally_defined_coord}
The radial coordinates do not have a chart coordinate subindex for they are globally defined by
$$r_{\varepsilon_j}(t)=f(x_{\varepsilon_j}(t)),\quad t\in\R,$$
$$r(t)=f(x(t)),\quad t\in [-T, T].$$
By Lemma \ref{Lema2}, these definitions are compatible with each other.
\end{rmk}

In what follows, we adopt the Einstein's summation convention on repeated indices. We denote by greek letters indices ranging from $0$ to $n-1$ and by $z$ the coordinates of the curves. We denote by latin letters indices ranging from $1$ to $n-1$ and by $y$ the coordinates of the curve. The zero index denotes the radial coordinate. In resume,
$$z^{0}= r,\ z^{1}= y^{1},\ldots,\ z^{n-1}=y^{n-1}.$$
For a better readability, we will denote the radial index zero by $r$.

Because of the construction of the coordinate charts, the pull-back of the metric $\rho$ by any of these has the form
$$\Psi^{*}(\rho)= \gi_{rr}\, dr\otimes dr + \gi_{ab}\, dy^{a}\otimes dy^{b}.$$
Note that there are no mixed indices terms or equivalently $\gi_{ra}={\bf 0}$. In particular, thinking of the metric as a matrix $(\gi_{\alpha\beta})$, its inverse $(\gi^{\alpha\beta})$ verifies
$$\gi^{rr}= \gi_{rr}^{-1},\quad (\gi^{ab})=(\gi_{ab})^{-1},\quad \gi^{ra}={\bf 0}.$$

The Christoffel symbols (of the second kind) are defined as follows:
$$\Gamma^{\alpha}_{\mu\nu}=\frac{\gi^{\alpha\beta}}{2}\left(\gi_{\beta\mu,\nu}+\gi_{\beta\nu,\mu}- \gi_{\mu\nu,\beta}\right)$$
where we have denoted by a comma the respective partial derivative. In contrast with the metric, now these symbols are not the coefficients of a tensor. However, with respect to the coordinate charts we are using, they are the coefficients of a connection, specifically, the Levi-Civita connection:
$$\nabla_{\partial_\mu}\partial_\nu= (\Gamma^{\alpha}_{\mu\nu}\circ \Psi^{-1})\ \partial_\alpha.$$

\begin{lema}\label{Lema4}
\begin{enumerate}
\item\label{item1_lema4} The functions $\dot{r}_{\varepsilon_j}$ and $\dot{y}_{i,\varepsilon_j}^{k}$ are uniformly bounded by a constant not depending on the chart index.
\item $r_{\varepsilon_j}\rightarrow {\bf 0}$ uniformly and $\dot{r}_{\varepsilon_j}\rightharpoonup {\bf 0}$ weakly star in $C[-T,T]$ as $j\to +\infty$.
\end{enumerate}
\end{lema}
\begin{proof}
\begin{enumerate}
\item For every $i=1,2,\ldots l$ define the function $Q_i$ on $T\,\R^{n}$ such that
$$Q_i({\rm x},{\rm v})= \Vert d_{{\rm x}} \Psi_i({\rm v})\Vert^{2},\qquad ({\rm x},{\rm v})\in T\,\R^{n}.$$
For every ${\rm x}$ it defines a positive definite quadratic form hence it defines a strictly positive continuous function on the unit tangent sphere bundle $\pi:T^{1}\R^{n}\to \R^{n}$. In particular, it attains a minimum value $m_i>0$ on the compact set $(\Psi_i\circ\pi)^{-1}(C_i)$. Define $m>0$ as the minimum of the $m_i$'s.

For every $\varepsilon>0$, $i=1,2,\ldots l$ and $t$ in $I_i$ we have
$$m\Vert(\dot{r}_{\varepsilon_j}(t), \dot{y}_{i,\varepsilon_j}(t))\Vert^{2}\leq 
Q_i\left((r_{\varepsilon_j}(t), y_{i,\varepsilon_j}(t)), (\dot{r}_{\varepsilon_j}(t), \dot{y}_{i,\varepsilon_j}(t))\right)$$
$$= \Vert \dot{x}_{\varepsilon_j}(t)\Vert_{x_{\varepsilon_j}(t)}^{2}\leq \Vert v \Vert_p^{2}$$
hence we conclude that
$$|\dot{r}_{\varepsilon_j}(t)|,\ |\dot{y}_{i,\varepsilon_j}^{k}(t))|\leq m^{-1/2}\Vert v \Vert_p$$
and the result follows.

\item Because of the uniform limit on $[-T,T]$
$$g(r_{\varepsilon_j})= U(x_{\varepsilon_j})\leq \varepsilon_j^{2}\ \Vert v \Vert_p^{2}/2\rightarrow 0,\qquad j\to +\infty$$
we have the first assertion. By the previous item, Proposition \ref{magic} and the fact that $C[-T,T]$ is a Banach subalgebra of $L^{\infty}[-T,T]$, we have the second assertion and the proof is complete.
\end{enumerate}
\end{proof}

\begin{lema}\label{Lema5}
For every natural $j$ and chart index $i$, considering the functions $r_{\varepsilon_j}$ and $\dot{r}_{\varepsilon_j}$ as external parameters, we have the nonautonomous equations
\begin{equation}\label{Equation_y}
\ddot{y}^{k}_{i,\varepsilon_j}+\Gamma^{k}_{ab}\;\dot{y}^{a}_{i,\varepsilon_j}\,\dot{y}^{b}_{i,\varepsilon_j}+
2\,\Gamma^{k}_{ar}\,\dot{y}^{a}_{i,\varepsilon_j}\,\dot{r}_{i,\varepsilon_j}+
\Gamma^{k}_{rr}\,\dot{r}^{2}_{i,\varepsilon_j}=0
\end{equation}
where the coefficients are evaluated on $y_{i,\varepsilon_j}$ and $r_{\varepsilon_j}$.
\end{lema}
\begin{proof}
With respect to the coordinate chart $(\I(V_i), \Psi_i)$, the Lagrangian \eref{Lagrangian} reads as follows:
$$L_\varepsilon\circ d_{({\rm r,y})}\Psi_i({\rm v})= \frac{\gi_{\alpha\beta}({\rm r,y})}{2}\,{\rm v}^{\alpha}{\rm v}^{\beta} -\varepsilon^{-2}g({\rm r}).$$
The Euler-Lagrange equations are covariant with respect to coordinate changes hence the coordinate curves $r_{\varepsilon_j}$ and $y_{i,\varepsilon_j}$ verify them. A straightforward calculation of the equations gives the result.
\end{proof}

Now, because of Lemma \ref{Lema4}, the motion equations \eref{Equation_y} are not singular as $j\to +\infty$.

\begin{cor}\label{Cor4}
There is a constant $A$ neither depending on $j$ nor on the chart index $i$ such that $\Vert\ddot{y}_{i,\varepsilon_j}(t)\Vert\leq A$ for every $j\geq N_0$, every $i=1,2,\ldots l$ and every $t$ in $I_i$.
\end{cor}
\begin{proof}
All of the coefficients in expression \eref{Equation_y} are continuous on $\Psi_i^{-1}(\I(V_i))$ hence they are bounded on the compact set $\Psi_i^{-1}(C_i)$ where they are evaluated. By Lemma \ref{Lema4}, all of the velocities are bounded by a constant not depending on $j$. Therefore, for every chart index $i$, by expression \eref{Equation_y} again the same occurs with the accelerations. Taking the maximum of these bounds with respect to the chart index $i$, we have the result.
\end{proof}

The following corollary is a standard argument and is the analog of Corollary 2.9 in \cite{BMP}.
\begin{cor}\label{Cor5}
For every chart index $i$, taking a subsequence if necessary, the tangential coordinates $y_{i,\varepsilon_j}$ converge strongly in the $C^{1}$ topology to $y_i$ on $\bar{I}_i\cap [-T, T]$. In particular, if zero is contained in $I_i$, then
$$d\psi_i \left(y_i(0),\dot{y}_i(0)\right)=(p,v_\Vert)$$
where $v_\Vert$ is the orthogonal projection of $v$ on $M$.
\end{cor}
\begin{proof}
Consider a chart index $i$. By Lemma \ref{Lema4}, the sequence $(\dot{y}_{\varepsilon_j})$ on $\bar{I}_i\cap [-T, T]$ is uniformly bounded and by Corollary \ref{Cor4} it is equicontinuous. By Arzel\`a--Ascoli Theorem, taking a subsequence if necessary, there is a continuous function $h$ such that $\dot{y}_{i,\varepsilon_j}\rightarrow h$ uniformly on $\bar{I}_i\cap [-T, T]$. We have
$$y_{i,\varepsilon_j}(t)= y_{i,\varepsilon_j}(\tau_i)+\int_{\tau_i}^{t}ds\ \dot{y}_{i,\varepsilon_j}(s),\quad t\in \bar{I}_i$$
and because $(y_{i,\varepsilon_j})$ and $(\dot{y}_{i,\varepsilon_j})$ converge uniformly on $\bar{I}_i\cap [-T, T]$ to $y_i$ and $h$ respectively, we have
$$y_i(t)= y_i(\tau_i)+\int_{\tau_i}^{t}ds\ h(s),\quad t\in \bar{I}_i\cap [-T, T]$$
and we conclude that $y_i$ is differentiable on $\bar{I}_i\cap [-T, T]$ and $\dot{y}_i=h$.

Finally, if zero is contained in $I_i$, then
$$(p,v_\Vert)= d\psi_i \left(y_{i,\varepsilon_j}(0),\dot{y}_{i,\varepsilon_j}(0)\right)\rightarrow d\psi_i \left(y_i(0),\dot{y}_i(0)\right)$$
as $j\to +\infty$ and we have the result.
\end{proof}

\begin{lema}\label{Lema6}
The limit curve $x$ in Corollary \ref{Cor3} is $C^{1}$ with $\dot{x}(0)= v_{\Vert}$, where $v_\Vert$ is the orthogonal projection of $v$ on $M$. Moreover, taking a subsequence if necessary, $y_{i,\varepsilon_j}$ converge strongly in the $C^{1}$ topology to $y_i$ on every $\bar{I}_i\cap [-T, T]$ \footnote{In contrast to Corollary \ref{Cor5}, now the same subsequence works for every coordinate chart.}.
\end{lema}
\begin{proof}
By Corollary \ref{Cor5}, there is a subsequence $(\varepsilon_j^{(1)})$ of $(\varepsilon_j)$ such that $y_{1,\varepsilon_j^{(1)}}$ converges strongly in the $C^{1}$ topology to $y_1$ on $\bar{I}_1\cap [-T, T]$. Inductively, there is a subsequence $(\varepsilon_j^{(i)})$ of $(\varepsilon_j^{(i-1)})$ such that $y_{i,\varepsilon_j^{(i)}}$ converges strongly in the $C^{1}$ topology to $y_i$ on $\bar{I}_i\cap [-T, T]$.
The subsequence $(\varepsilon_j^{(l)})$ is the one we were looking for. Now, for every chart index $i$, $y_{i,\varepsilon_j^{(l)}}$ converges strongly in the $C^{1}$ topology to $y_i$ on every $\bar{I}_i\cap [-T, T]$.

In particular, for every chart index $i$, $y_i$ is $C^{1}$ on every $\bar{I}_i\cap [-T, T]$ and because the open intervals $I_1,\ldots I_l$ cover $[-T,T]$, we conclude that $x$ is $C^{1}$.

Finally, if zero is contained in $I_i$, then
$$v_\Vert= d_{y_i(0)}\psi_i \left(\dot{y}_i(0)\right)= \dot{x}(0)$$
for $x|_{\bar{I}_i}= \psi_i\circ y_i$ and we have the result.
\end{proof}

From now on, we will consider the subsequence of $(\varepsilon_j)$ starting from $N_0$. In subsection \ref{Subsection_proof}, it will be shown that the limit curve $x$ is actually $C^{2}$ and the whole family $(x_\varepsilon)_{\varepsilon>0}$ uniformly converges to it on $[-T, T]$.


\subsection{The radial adiabatic invariant}

To simplify the notation, define $h_r(x)= \Vert\grad_\rho f(x)\Vert^{-1}$. It is the radial scale factor with respect to any of the coordinate charts considered before:
\begin{equation}\label{square_radial_scalefactor}
h_r^{2}\circ \Psi= \gi_{rr}.
\end{equation}
However, in contrast to $\gi_{rr}$, it is globally defined and independent of these charts for the radial coordinate is globally defined by the function $f$.

The proof of the following lemma is an adaptation of an argument in (\cite{Bornemann}, chapter 1, sections 2.3 and 2.4).

\begin{lema}\label{Lema7}
Taking a subsequence if necessary, the following sequences weakly converge in $C[-T, T]$:
$$\varepsilon_j^{-2}g(r_{\varepsilon_j})\rightharpoonup \sigma,\quad \quad \dot{r}_{\varepsilon_j}^{2}\rightharpoonup \pi$$
where $\sigma$ and $\pi$ are $C^{1}$ functions on $[-T,T]$ verifying the following relations:
\begin{enumerate}
\item $\alpha\ h_r(x)^{2}\,\pi =  \sigma$ on $[-T, T]$.
\item There is a constant $\theta$ such that $h_r(x)^{2+\frac{2}{2\alpha+1}}\ \pi=\theta$ on $[-T, T]$.
\end{enumerate}
\end{lema}
\begin{proof}
Since the sequences $\varepsilon_j^{-2}g(r_{\varepsilon_j})$ and $\dot{r}_{\varepsilon_j}^{2}$ are uniformly bounded by item \ref{item2_lema_prel} in Lemma  \ref{lema_preliminaries} and item \ref{item1_lema4} in Lemma \ref{Lema4} respectively, by Proposition \ref{Alaoglu} and taking a subsequence if necessary, there are functions $\sigma$ and $\pi$ in $L^{\infty}[-T,T]$ such that the sequences
\begin{equation}\label{sigma_pi}
\varepsilon_j^{-2}g(r_{\varepsilon_j})\rightharpoonup \sigma,\quad \quad \dot{r}_{\varepsilon_j}^{2}\rightharpoonup \pi
\end{equation}
weakly star converge in $L^{\infty}[-T, T]$.

\begin{enumerate}
\item
Denote by $e$ the resulting $C^{1}$ function on $I$ after removing the singularity of $g/g'$ at the origin by defining $e(0)=0$. Consider the subsequence $(\varepsilon_j)_{j\geq j_0}$ with $j_0$ big enough such that the functions $r_{\varepsilon_j}$ are $I$ valued on $[-T,T]$ for every $j\geq j_0$. Recall that $e'(0)=\alpha$.

Consider an arbitrary coordinate chart $(\I(V_i), \Psi_i)$. With respect to this chart, a verbatim argument as in Lemma \ref{Lema5} shows that the radial motion equation reads as follows:
\begin{equation}\label{Equation_r}
\ddot{r}_{\varepsilon_j}+\Gamma^{k}_{\alpha\beta}\;\dot{z}^{\alpha}_{i,\varepsilon_j}\,\dot{z}^{\beta}_{i,\varepsilon_j}
+\varepsilon_j^{-2} h_r^{-2}(x_{\varepsilon_j})g'(r_{\varepsilon_j})=0.
\end{equation}
All of the coefficients are continuous on $\Psi_i^{-1}(\I(V_i))$ hence they are bounded on the compact set $\Psi^{-1}(C_i)$ where the coordinate curves live and the coefficients are evaluated. The velocities are uniformly bounded hence the equation \eref{Equation_r} has the form
$$\ddot{r}_{\varepsilon_j}= b_{i,\varepsilon_j} - \varepsilon_j^{-2} h_r^{-2}(x_{\varepsilon_j})g'(r_{\varepsilon_j})$$
where the functions $b_{i,\varepsilon_j}$ are uniformly bounded. In particular,
$$e(r_{\varepsilon_j})\ddot{r}_{\varepsilon_j}= e(r_{\varepsilon_j})b_{i,\varepsilon_j} - h_r^{-2}(x_{\varepsilon_j})\left(\varepsilon_j^{-2}g'(r_{\varepsilon_j})e(r_{\varepsilon_j})\right)$$
\begin{equation}\label{convergence1}
=e(r_{\varepsilon_j})b_{i,\varepsilon_j} - h_r^{-2}(x_{\varepsilon_j})\left(\varepsilon_j^{-2}g(r_{\varepsilon_j})\right)
\rightharpoonup - h_r^{-2}(x) \sigma
\end{equation}
for the first term on the r.h.s. uniformly goes to zero and we have used Proposition \ref{module} on the second term. Note that the limit \eref{convergence1} is independent of the chart index and because it was arbitrary, this limit holds on the entire interval $[-T, T]$.

Define the auxiliar function $c_{\varepsilon_j}= e(r_{\varepsilon_j})\dot{r}_{\varepsilon_j}$ and see that it uniformly goes to zero as $j\to+\infty$. Because of expressions \eref{sigma_pi} and \eref{convergence1}, its derivative has the weak star limit
$$\dot{c}_{\varepsilon_j}= e(r_{\varepsilon_j})\ddot{r}_{\varepsilon_j}+ e'(r_{\varepsilon_j})\dot{r}_{\varepsilon_j}^{2}\rightharpoonup - h_r^{-2}(x) \sigma +\alpha\pi.$$
where we have used Proposition \ref{module} on the second term. In particular, by Proposition \ref{Bound}, these derivatives are uniformly bounded and because $c_{\varepsilon_j}$ uniformly goes to zero, by Proposition \ref{magic} we have $\dot{c}_{\varepsilon_j}\rightharpoonup 0$ weakly star in $L^{\infty}[-T,T]$. The uniqueness of the limit implies the first relation.

\item Define the radial kinetic and potential energy respectively by
$$T_{\varepsilon_j}^{\perp}= \frac{h_r(x_{\varepsilon_j})^{2}}{2}\dot{r}_{\varepsilon_j}^{2},\quad \quad U_{\varepsilon_j}^{\perp}=\varepsilon_j^{-2}g(r_{\varepsilon_j}).$$
Define the total radial energy as their sum $E_{\varepsilon_j}^{\perp}= T_{\varepsilon_j}^{\perp}+U_{\varepsilon_j}^{\perp}$. By Proposition \ref{module} and expression \eref{sigma_pi},
\begin{equation}\label{one_hand}
E_{\varepsilon_j}^{\perp}\rightharpoonup E_0^{\perp}= \frac{h_r(x)^{2}}{2}\pi+ \sigma= (2\alpha+1)\frac{h_r(x)^{2}}{2}\pi
\end{equation}
weakly star in $L^{\infty}[-T,T]$ where we have used the first relation.

Now we calculate the derivative of the total radial energy. With respect to some coordinate chart $(\I(V_i), \Psi_i)$ and recalling expression \eref{square_radial_scalefactor} we have
$$\dot{E}_{\varepsilon_j}^{\perp}= \gi_{rr}(z_{i,\varepsilon_j})\dot{r}_{\varepsilon_j}\ddot{r}_{\varepsilon_j}+ \frac{\dot{r}_{\varepsilon_j}^{2}}{2}\gi_{rr,\alpha}(z_{i,\varepsilon_j})\dot{z}_{i,\varepsilon_j}^{\alpha} + \varepsilon_j^{-2}g'(r_{\varepsilon_j})\dot{r}_{\varepsilon_j}$$
$$=\dot{r}_{\varepsilon_j}\left(\frac{d}{d\tau}\frac{\partial L_{\varepsilon_j}}{\partial v_r}-\frac{\partial L_{\varepsilon_j}}{\partial r}\right)(z_{i,\varepsilon_j}, \dot{z}_{i,\varepsilon_j})+\frac{\gi_{ab,r}(z_{i,\varepsilon_j})}{2}\dot{y}_{i,\varepsilon_j}^{a}\dot{y}_{i,\varepsilon_j}^{b}\dot{r}_{ \varepsilon_j}$$
$$-\frac{\dot{r}_{\varepsilon_j}^{2}}{2}\gi_{rr,a}(z_{i,\varepsilon_j})\dot{y}_{i,\varepsilon_j}^{a}$$
\begin{equation}\label{other_hand}
=\frac{\gi_{ab,r}(z_{i,\varepsilon_j})}{2}\dot{y}_{i,\varepsilon_j}^{a}\dot{y}_{i,\varepsilon_j}^{b}\dot{r}_{\varepsilon_j}
-\frac{\dot{r}_{\varepsilon_j}^{2}}{2}\gi_{rr,a}(z_{i,\varepsilon_j})\dot{y}_{i,\varepsilon_j}^{a}.
\end{equation}
 
Then, by Propositions \ref{Bound}, \ref{compact_sets} and Lemma \ref{Lema4}, the radial energy sequence $E_{\varepsilon_j}^{\perp}$ belongs to and is uniformly bounded in $W^{1,\infty}(-T,T)$. By Proposition \ref{Alaoglu_W}, there is a subsequence of it that weakly star converges in $W^{1,\infty}(-T,T)$ to some $E_0^{\perp}$ in this space. In particular, by equation \eref{one_hand}, the function $\pi$ is also in $W^{1,\infty}(-T,T)$ hence it has a weak derivative in $L^{\infty}(-T,T)$ and we have the equation
\begin{equation}\label{eq2}
\dot{E}_0^{\perp}= (2\alpha+1)\frac{d}{dt}\left(\frac{h_r(x)^{2}}{2}\pi\right).
\end{equation}
On the other hand, because of the expression \eref{other_hand} and the limit uniqueness we have
\begin{equation}\label{eq3}
\dot{E}_{\varepsilon_j}^{\perp}\rightharpoonup  \dot{E}_0^{\perp} = -\frac{\pi}{2}\gi_{rr,a}(0,y_i)\dot{y}_i^{a}
=-\frac{\pi}{2}\frac{d}{dt}\left(h_r(x)^{2}\right).
\end{equation}
where we have used expression \eref{square_radial_scalefactor} again. By the expressions \eref{eq2} and \eref{eq3} we conclude that $\pi$ is a weak solution in $W^{1,\infty}(-T,T)$ of the ordinary differential equation
\begin{equation}\label{diff_eq_pi}
\dot{\pi}= -\left(1+\frac{1}{2\alpha+1}\right)\ \frac{d \log h_r(x)^{2}}{dt} \pi.
\end{equation}

Now, this equation is independent of the chart index and because it was arbitrary, the equation holds on the entire interval $[-T,T]$. By Proposition \ref{weak_strong}, the a priori weak solution $\pi$ is actually a strong solution in $C^{1}[-T,T]$ and unique after an initial condition is given. In particular, there is a constant $\theta$ such that
\begin{equation}\label{pi_resultado}
\pi= \theta h_r(x)^{-2-\frac{2}{2\alpha+1}}
\end{equation}
on $[-T, T]$. This proves the second relation.
\end{enumerate}
In particular, because $C[-T,T]$ is a Banach subalgebra of $L^{\infty}[-T,T]$\footnote{This is the reason for the choice of this particular $L^{p}$ space in the proof.}, the weak star convergence to $\pi$ is actually in $C[-T,T]$ and because of the first relation the same holds for $\sigma$. The proof is complete.
\end{proof}

\begin{cor}\label{Adiabatic_Invariant}
$$\theta\,=\,\frac{1}{2\alpha+1}\, \Vert v_{\perp}\Vert_p^{2}\ \Vert\grad_\rho f(p)\Vert_p^{-\frac{2}{2\alpha+ 1}}.$$
\end{cor}
\begin{proof}
By the weak convergence \eref{one_hand} and expression \eref{pi_resultado} in the proof of the previous lemma, the total energy weakly converges to
$$E_{\varepsilon_j}\rightharpoonup \frac{\gi_{ab}(0,y_i)}{2}\,\dot{y}^{a}_i\,\dot{y}^{a}_i\,+\,
(2\alpha+1)\,\frac{\theta}{2}\, h_r(x)^{-\frac{2}{2\alpha+1}},$$
for every chart index $i$, where we have used the strong convergence in Lemma \ref{Lema4} and \ref{Lema6}. However, the total energy is independent of $j$ and equals
$$E_{\varepsilon_j}\,=\,\frac{h_r(p)^2}{2}\,\dot{r}_0^2
\,+\,\frac{\gi_{ab}(0,y_i(0))}{2}\,\dot{y}^{a}_0\,\dot{y}^{a}_0,$$
where we have restricted to the chart index $i$ such that the interval $I_i$ contains zero (see Corollary \ref{Cor5}) and we have denoted with a zero subindex the coordinates of the initial velocities.

By the uniqueness of the weak limit, the right hand side of both expressions are equal hence evaluating the right hand side of the first expression at time zero we have
$$\frac{h_r(p)^2}{2}\,\dot{r}_0^2\,=\,(2\alpha+1)\,\frac{\theta}{2}\, h_r(p)^{-\frac{2}{2\alpha+1}}.$$
Solving for $\theta$ and recalling the definitions, we have the result.
\end{proof}

If we express the \textit{adiabatic invariant} $\theta$ in terms of the coordinate of the radial velocity $v_r$ instead of its norm $\Vert v_{\perp}\Vert_p$ as we did in the previous corollary, then we get the same result as in (Example 1, \cite{Bornemann}, page 24.) for the codimension one nondegenerate case, i.e. $\alpha=1/2$.

\subsection{Equipotential distortion}\label{Subsection_eq}

Consider an arbitrary chart index and its associated coordinate chart $(\I(V),\Psi)$. With respect to this chart, the coordinates of the equipotential distortion $\kappa$ read as follows
\begin{equation}\label{coord_kappa}
\kappa_a= \left\langle \kappa\circ\Psi, \partial_a\Psi \right\rangle,\qquad \kappa^{a}= \gi^{ab}\,\kappa_b.
\end{equation}

\begin{lema}\label{Lema8}
With respect to the coordinate chart $(\I(V),\Psi)$, we have the relations
$$\Gamma_{rr}^{a}= \gi_{rr}\,\kappa^{a},\qquad \kappa_a= -\frac{1}{2}\partial_a \log \gi_{rr}.$$
\end{lema}
\begin{proof}
Define the radial scale factor $s_r$ and unit vector $e_r$ such that $\partial_r\Psi= s_r\, e_r$. Equivalently, $s_r= h_r\circ\Psi$. By definition of the equipotential distortion,
$$\kappa\circ\Psi= s_r^{-1}\nabla_r e_r$$
where $\nabla$ is the Levi-Civita connection with respect to the metric $\rho$. By definition of the coordinates we have
$$\left\langle\partial_a\Psi, e_r\right\rangle\equiv 0.$$
Because the Levi-Civita connection has no torsion, we have
$$\nabla_a(\partial_r\Psi)-\nabla_r(\partial_a\Psi)= [\partial_a\Psi, \partial_r\Psi]\equiv 0.$$
The second relation follows immediately from
$$\kappa_a= \left\langle\partial_a\Psi, \kappa\circ\Psi\right\rangle= \left\langle\partial_a\Psi, s_r^{-1}\nabla_r e_r \right\rangle
=-s_r^{-1}\left\langle\nabla_r(\partial_a\Psi),  e_r \right\rangle$$
$$= -s_r^{-1}\left\langle\nabla_a(\partial_r\Psi),  e_r \right\rangle= -s_r^{-2}\left\langle\nabla_a(\partial_r\Psi), \partial_r\Psi  \right\rangle
= -\gi_{rr}^{-1}\,\gi_{rr,a}/2$$
while the first relation follows from
$$\gi_{rr}\,\kappa^{a}= -\gi^{ab}\,\gi_{rr,b}/2= \Gamma_{rr}^{a}.$$
\end{proof}

\begin{cor}\label{Cor6}
We have the following relation on $N-Crit(f)$:
$$\kappa=\left(\grad_\rho\log\Vert\grad_\rho f\Vert\right)_{\Vert}$$
where the subsymbol $\Vert$ denotes the orthogonal projection on the respective equipotential. In particular, given an equipotential hypersurface, its equipotential distortion is null iff $\Vert\grad_\rho f\Vert$ is constant on it.
\end{cor}
\begin{proof}
For every coordinate chart $(\I(V),\Psi)$ we have $\gi_{rr}= \Vert\grad_\rho f\Vert^{-2}\circ\Psi$ and the result follows from the definition \eref{coord_kappa} and the expression
$$\kappa_{a}= -\frac{1}{2}\partial_a \log \gi_{rr}= \partial_a (\log \Vert\grad_\rho f\Vert\circ\Psi)=
\left\langle \left(\grad_\rho\log\Vert\grad_\rho f\Vert\right)\circ\Psi, \partial_a\Psi \right\rangle$$
for every tangential index $a$.
\end{proof}

\subsection{Proofs}\label{Subsection_proof}

\begin{proof}[Proof of Theorem \ref{Main}]
Consider an arbitrary chart index and its associated coordinate chart $(\I(V_i),\Psi_i)$. With respect to this chart, writing equations \eref{Equation_y} in integral form we have
$$\dot{y}_{i,\varepsilon_j}^{k}(t)= \dot{y}_{i,\varepsilon_j}^{k}(\tau_i)-\int_{\tau_i}^{t}ds\ \left(
\Gamma^{k}_{ab}\,\dot{y}^{a}_{i,\varepsilon_j}\,\dot{y}^{b}_{i,\varepsilon_j}+
2\,\Gamma^{k}_{ar}\,\dot{y}^{a}_{i,\varepsilon_j}\,\dot{r}_{\varepsilon_j}+
\Gamma^{k}_{rr}\,\dot{r}^{2}_{\varepsilon_j}\right).$$
Taking the limit as $j\to+\infty$,
$$\dot{y}_i^{k}(t)= \dot{y}_i^{k}(\tau_i)-\int_{\tau_i}^{t}ds\ \left(
\Gamma^{k}_{ab}\,\dot{y}_i^{a}\,\dot{y}_i^{b}+
\Gamma_{rr}^{k}\,\pi \right).$$
By Lemmas \ref{Lema7} and \ref{Lema8}, we have
\begin{equation}\label{Integral_eq}
\dot{y}_i^{k}(t)= \dot{y}_i^{k}(\tau_i)-\int_{\tau_i}^{t}ds\ \left(
\Gamma^{k}_{ab}\,\dot{y}_i^{a}\,\dot{y}_i^{b}+
\theta\ \gi_{rr}^{-\frac{1}{2\alpha+ 1}}\, \kappa^{k} \right).
\end{equation}
where $\theta$ is the adiabatic invariant given in Corollary \ref{Adiabatic_Invariant}.

Now, independently of any coordinate chart, from equations \eref{Integral_eq} it immediately follows that the limit curve $x$ is at least $C^{2}$  and it is the unique solution of
\begin{equation}\label{Intrinsic_eq}
\nabla_{\dot{x}}\,\dot{x} + \theta\ \Vert\grad_\rho f(x)\Vert_x^{\frac{2}{2\alpha+ 1}}\ \kappa(x)=0,\quad x(0)=p,\quad \dot{x}(0)= v_\Vert
\end{equation}
where $\nabla$ denotes the Levi-Civita connection on $M$ with respect to the induced ambient metric.

We have proved that every sequence $(x_{\varepsilon_n})$ has a uniformly convergent subsequence to the unique solution $x$ of the Cauchy problem \eref{Intrinsic_eq}. We conclude that the whole family $(x_\varepsilon)_{\varepsilon>0}$ uniformly converges on $[-T, T]$ to $x$ for otherwise, there would be a sequence $(x_{\varepsilon_n})$ whose distance to $x$ is greater than or equal to some $\delta>0$ and containing a uniformly convergent subsequence to $x$ which is absurd.

Because the choice of $T>0$ was arbitrary, we have the result.
\end{proof}

\begin{proof}[Proof of Corollary \ref{Main_Cor_1}]
By Corollary \ref{Cor6}, we have
$$\theta\ \Vert\grad_\rho f\Vert^{\frac{2}{2\alpha+ 1}}\ \kappa= \theta\ \Vert\grad_\rho f\Vert^{\frac{2}{2\alpha+ 1}}
\left(\grad_\rho\log\Vert\grad_\rho f\Vert\right)_{\Vert}$$
\begin{equation}\label{gradient_orthoginal}
=\theta\ \Vert\grad_\rho f\Vert^{\frac{2}{2\alpha+ 1}-1}
\left(\grad_\rho\Vert\grad_\rho f\Vert\right)_{\Vert}= \left(\grad_\rho\ U_{eff}\right)_\Vert.
\end{equation}
where $\theta$ is the adiabatic invariant given in Corollary \ref{Adiabatic_Invariant}. Because the orthogonal projection on $M$ of the gradient with respect to the metric $\rho$ is the gradient with respect to the induced ambient metric on $M$, we have that the equation \eref{Intrinsic_eq} coincides with the Euler-Lagrange equations of the Lagrangian $L_{eff}$ on $TM$ and the proof is complete.
\end{proof}

Corollaries \ref{Main_Cor_2} and \ref{Main_Cor_3} are immediate from Theorem \ref{Main}.

\ack

The author is a researcher at \textit{Consejo Nacional de Ciencia y Tecnolog\'ia, CONACYT}.

\References

\bibitem[Ad]{Adams}
Adams R A 1975, \emph{Sobolev Spaces}, Academic Press, New York.

\bibitem[Ar]{Arnold}
Arnold V I 1978, \emph{Mathematical methods of classical mechanics}, Springer-Verlag, Berlin, Heidelberg, New York.

\bibitem[AM]{AbrahamMarsden}
Abraham R, Marsden J E 1987, \emph{Foundations of Mechanics}, Addison-Wesley, Publishing Company, Redwood City, Second Edition.

\bibitem[Bo]{Bornemann}
Bornemann F A 1998, \emph{Homogenization in Time for Singularly Perturbed Mechanical Systems}, Lecture Notes in Mathematics {\bf 1687}, Springer.

\bibitem[BS]{BS}
Bornemann F A, Sch\"utte C 1997, \emph{Homogenization of Hamiltonian systems with a strong constraining potential}, Phys. D, {\bf 102}, 57--77.

\bibitem[BP]{BP}
Burgos J M, Paternain M 2022, \emph{On the Lyapunov instability in Lagrangian dynamics}, Proc. Amer. Math. Soc. {\bf 150}, 4335--4348.

\bibitem[BMP]{BMP}
Burgos J M, Maderna E, Paternain M 2021, \emph{On the Lyapunov instability in Newtonian dynamics}, Nonlinearity, {\bf 34}, 6719--6726. 

\bibitem[EZ]{Evans_av}
Evans L C, Zhang T 2016, \emph{Weak convergence and averaging for ODE}, Nonlinear Analysis: Theory, Methods and Applications {\bf 138}, 83--92.

\bibitem[La]{Lagrange}
Lagrange J L 1787, \emph{Mecanique Analytique}, Mme Ve Couricer, Paris.

\bibitem[Ga]{Gallavotti}
Gallavotti G 1983, \emph{The Elements of Mechanics}, Springer-Verlag, Berlin, Heidelberg, New York.

\bibitem[KJ]{KoppeJensen}
Koppe H, Jensen H 1971, \emph{Das Prinzip von d'Alembert in der Klassischen Mechanik und in der Quantentheorie}, Sitzungsberichte der Heidelberger Akademie der Wissenschaften {\bf 5}.

\bibitem[vK]{vanKampen}
van Kampen N G 1985, \emph{Elimination of fast variables}, Phys. Rep. {\bf 124} 69--160.

\bibitem[Ru]{Rudin}
Rudin W 1976, \emph{Principles of Mathematical Analysis}, Madison, WI.

\bibitem[RU]{RubinUngar}
Rubin H, Ungar P 1957, \emph{Motion under a strong constraining force}, Comm. Pure. Applied Math., {\bf 10}, 65--87.

\bibitem[Ta]{Takens}
Takens F 1980, \emph{Motion under the influence of a strong constraining potential}, Global Theory of Dynamical Systems, Z. Nitecki and C. Robinson eds., Springer-Verlag, Berlin, Heidelberg, New York, 425--445.

\endrefs

\end{document}